 \documentclass[final,5p,times,twocolumn,authoryear]{elsarticle}

\usepackage{amssymb}
\usepackage{lipsum}

\usepackage{xcolor}

\usepackage[colorlinks=true, allcolors=blue]{hyperref}
\usepackage{amsmath}
\usepackage{booktabs}
\usepackage{multirow}
\usepackage{orcidlink}

\journal{High Energy Astrophysics}

\begin{document}

\begin{frontmatter}



\title{Particle acceleration to PeV energies in Pulsar Wind Nebula: a two zone model}


\author[first,second]{Gunindra Krishna Mahanta \, \orcidlink{0000-0003-4544-2585}}
\ead{guninmohantaba@gmail.com}
\author[first]{Nilay Bhatt}
\author[first,second]{Bitan Ghosal}
\author[first,second]{Subir Bhattacharyya}
\ead{subirb@barc.gov.in}
\affiliation[first]{organization={Astrophysical Sciences Division},
            addressline={Bhabha Atomic Research Centre}, 
            city={Mumbai},
            postcode={400085}, 
            state={Maharashtra},
            country={India}}

\affiliation[second]{organization={Homi Bhabha National Institute},
            addressline={Anushaktinagar}, 
            city={Mumbai},
            postcode={400094}, 
            state={Maharashtra},
            country={India}}

\begin{abstract}
\textit{PeVatrons} are the extreme galactic accelerators capable of producing PeV particles. Recent observation of Large High Altitude Air Shower Observatory have detected UHE photons ($\geq$ 100 TeV) from 43 galactic sources. Detection of UHE photons demands the presence of at least PeV particles in the acceleration site. Although the exact nature of most of the sources are still unknown, a large fraction of these sources have spatial association with pulsar wind nebula. In this work we investigate the acceleration mechanism in pulsar wind nebula by following a magnetohydrodynamics approach. Current study relates the MHD flow solution in immediate downstream with the particle spectrum and spectral energy distribution of photons. Our study shows that MHD description in the PWN environment reduces the parameter space and most of the parameters can be constrained in terms of a single parameter, \textit{the magnetization parameter} $\sigma$ only. Considering the effect of $\sigma$, we show that in low $\sigma$ environment pulsar wind nebula can produce PeV particles. We have also investigate the role of turbulence in the nebular region in acceleration of particle to  PeV energy. Current study shows that both low $\sigma$ environment and turbulence environment is favorable for acceleration of particles up to PeV energy. We have also tested our model in two different LHAASO detected PeVatron 1LHAASO J1848-000u and 1LHAASO J1929+1846u.
\end{abstract}




\end{frontmatter}




\section{Introduction}
\label{sec:intro}

\noindent The search of origin of cosmic ray (CR) is one of the prime interests in high energy astrophysics. It is observed that CR spectrum from energy 10 GeV to $10^{11}$ GeV can be expressed as a power law with two spectral breaks around 3 PeV and 4 EeV \citep{ANTONI_2005,Thoudam_2016,Kang_2024}. At the first spectral break (occurs around 3 PeV), spectral index changes from -2.7 to -3.1, commonly referred as \textit{knee} of the CR spectrum, and the second spectral break (occurs around 4 EeV) CR spectrum flattening back to spectral index -2.7, commonly referred as \textit{ankle} of the CR spectrum \citep{Thoudam_2016}. It is generally argued that CRs particle up to the first spectral break so called \textit{knee} of the CR spectrum originate in our own galaxy \citep{Ptuskin_1993,Horandel_2003}, whereas CR particles beyond the ankle are argued to have extra galactic origin \citep{Rachen_1993,Berezinsky_2004}. Therefore the astrophysical sources which can accelerate particle up to PeV energy are the sources of special interest as they are extreme galactic accelerators. The sources which can produce PeV particles are called \textit{PeVatrons}. The PeV particles produced in the accelerator site can interact with its environment and produce $\gamma$-rays with energy $\geq$ 0.1 PeV (100 TeV). So the signature of both hadronic and leptonic PeVatron is the emission of photons with energy $\geq$ 100 TeV. In the recent decade emission of photons with energy $\geq$ 100 TeV is reported by various telescopes such as Tibet AS$\gamma$, HAWC, MAGIC etc. \citep{Amenomori_2019,Abeysekara_2020, Albert_2020,MAGIC_2020}. But the era of PeVatron physics began with the detection of 12 galactic PeVatron sources by Large High Altitude Air Shower Observatory (LHAASO) \citep{Cao_2021}. Out of this one candidate is Crab nebula, which is a pulsar wind nebula. Recently, LHAASO have released its 1$^\mathrm{st}$ catalog, in which UHE emission ($\geq$ 100 TeV) is reported from 43 galactic sources \citep{Cao_2024}. 

\noindent Diffusive shock acceleration (DSA) in supernova remnant (SNR) was considered to be primary acceleration process of galactic CRs (e.g. \cite{Blandford_1987}). But the maximum achievable energy through DSA in SNR shock is far below $\sim$ PeV. When instabilities like resonant streaming instability is considered, the maximum achievable energy comes out to be $\sim$ 0.1 PeV, which is also one order magnitude less than the maximum achievable energy in our galaxy \citep{Kulsrud_1969,Tatischeff_2018,Gabici_2019}. Even with non-resonant streaming instabilities, like Bell instability \citep{Bell_2004}, although it is possible to achieve instantaneous maximum energy $>$ 1 PeV, but when integrated over the history of SNR, the typical maximum energy released by SNR to ISM lies somewhere between 10 - 100 TeV \citep{Bell_2013, Cardillo_2016,Cristofari_2020}. These observations indicate that in addition to SNR, other class of galactic sources such as Pulsar Wind Nebulae, Young Massive Clusters inside superbubbles, Microquasars and Pulsars might be responsible for production of galactic CRs up to the knee \citep{Cristofari_2021}. It should be noted that, in addition to these candidates, molecular clouds (MCs) are also potential sites of ultra-high-energy emission. In systems involving supernova remnant–molecular cloud (SNR–MC) interactions, the high-energy emission can appear with a delay relative to the supernova explosion. This delay arises from the energy dependence of the particle diffusion coefficient, which allows the highest-energy particles to escape the remnant first and reach the surrounding molecular clouds, where they subsequently interact with dense matter. These interactions produce neutral pions through proton–proton collisions, whose decay leads to delayed ultra-high-energy gamma-ray emission \citep{Celli_2019a,Celli_2019b,Celli_2020}.

\noindent Apart from these theoretical arguments there are also several observational evidences which hint towards the scenario that SNRs are very less likely to be a PeVatron candidate. In the LHAASO 1$^\mathrm{st}$ catalog there is very few sources which have spatial association with SNR. Out of the 43 LHAASO detected UHE sources, 22 sources have spatial association with pulsar and among them 12 sources have spatial association with Pulsar Wind Nebula (PWN), some of them have spatial association with star cluster like Cygnus Cocoon, rest are unidentified \citep{Cao_2024}. Based on the observed data on LHAASO 1$^\mathrm{st}$ catalog it seems that PWN can also be a potential candidate to accelerate particle up to PeV energies. Previously observation of UHE ($>$ 100 TeV) emission was considered as a strong evidence of hadronic accelerator, as electron suffers Klein-Nishina suppression at UHE range. However after observations of 100 TeV $\gamma$-ray emission from PWN environment, which is considered as a standard electron accelerator, possibility of UHE electron accelerator opens. Later, by considering a one zone model \citet{Breuhaus_2022} shows that Inverse Compton scattering of electron can naturally explain the observed UHE emission from PWN environment. \\
In this work we formulate a self-consistent model to explain the observed UHE emission from the LHAASO detected PeVatron in the framework of pulsar wind nebula. The content of this manuscript is arranged as follows: In section \ref{sec:pwn}, we have  briefly discussed the current status and understanding of PWN, in section \ref{sec:model}, we presented the formulation of our model, including the treatment of particle acceleration at the termination shock, and the modeling of turbulent acceleration, radiative losses, and the time evolution of particles in the nebular environment, in section \ref{sec:Results} we discussed the effect of various parameter of our model in the particle spectrum and spectral energy distribution of photons emitted by the PeVatron sources, in section \ref{sec:application}, we discussed the application of our model in two different LHAASO detected PeVatron which have spatial association with PWN, finally in section \ref{sec:conclusion} we summarized the various aspect of our model.

\section{Current understanding of pulsar wind nebula}
\label{sec:pwn}

\noindent Pulsars are the rapidly rotating compact objects formed as a result of core collapse supernova explosion. The initial spin of pulsar gradually slows down and it releases its rotational energy through ultra-relativistic flow of electron, positron and magnetic energy commonly referred as \textit{pulsar wind}. Initially the pulsar wind flows at a bulk ultra-relativistic speed. This ultra-relativistic flow of electron-positron wind slows down due to the external pressure and finally halts when the wind pressure is balanced by the external pressure \citep{Vink_2020_book}, and create a shock called `\textit{termination shock}'. The region outside the termination shock is commonly referred to as Pulsar Wind Nebula (PWN) which shines in entire range of electromagnetic spectrum (ranging from radio to $\gamma$-ray) through synchrotron radiation and inverse Compton scattering. \\
The dynamics and evolution of PWN depends on the properties of the pulsar (eg. period, period derivative, initial spin down luminosity, breaking index etc.) as well as on the properties of the progenitor supernova (e.g. mass of the ejecta, explosion energy etc.) \citep{Gelfand_2009}. Evolution of the PWN inside the supernova remnant (SNR) is complex and different for different PWN. Still, in general the evolution of PWN inside a SNR can be categorize in three different phases \citep{Gaensler_2006}. First phase is called `free expansion phase' in which PWN expand freely with the SNR, in this phase expansion velocity of the PWN is much higher than velocity of the pulsar so that PWN can expand freely inside the SNR along with the pulsar. This phase ends when PWN collides with SNR reverse shock. The second phase of evolution commonly referred as `Re-vibration phase' begins after the first collision of PWN outer layer with the SNR reverse shock. At this phase PWN can not freely expand inside the SNR cell. Since initially pressure of SNR reverse shock (RS) is greater than the PWN pressure, so PWN start contract. As a result of compression, internal pressure of the PWN raises adiabatically, and eventually become higher than its surrounding pressure. The rapid increase in internal pressure cause the PWN to expand inside the cell again. This expansion-compression phase continues until the system reach some sort of thermal equilibrium.\\
During this re-vibration phase, it may happen that at some point of time expansion velocity of PWN become smaller than the pulsar kick velocity. At this point pulsar can detach itself from the PWN. \\
The last phase of evolution is commonly referred to as `relic PWN phase'. This phase starts when the pulsar detach itself from the relic PWN, and it forms a new PWN with relativistic positronic wind. The PWN around the neutron star and relic PWN evolve differently. The relic PWN will continue its re-vibration phase until it achieve pressure equilibrium with its surrounding.

\noindent Several attempt have made to understand the radiative properties and dynamical evolution of PWN. Initial hydrodynamic description of pulsar-wind nebula environment was outlined by \cite{Rees_1974}. \cite{Kennel_Coroniti_1984,Kennel_1984} gives the first magnetohydrodynamics (MHD) description of pulsar wind nebula . Evolution of PWN inside the SNR shell is also modeled in various literature \citep{Ostriker_1971,Pacini_1973,Chevalier_1982,Chevalier_1992,Blondin_1998,Van_2001,Bucciantini_2003, Bucciantini_2004a, Bucciantini_2004B,  Chevalier_2005, Gelfand_2007,Volpi_2008, Zhu_2023}. \cite{Gelfand_2009} gives a model which can explain both dynamical and radiative evolution of PWN  throughout its lifetime. Effect of turbulence in PWN is discussed in \cite{Lu_2023}. Apart from the radiative and dynamical modeling, recently PWN as a PeVatron candidate is also reported in literature. By considering time evolution of PWN \cite{Zhu_2024} explain the observed UHE emission as a result of inverse Compton scattering of UHE electron with star light, CMB, IR photons in the framework of PeVatron  source HESS J1848-000. \cite{Arons_1979} argued the observed PeV emission from crab as a result of reconnection in the persistent current layer. By introducing a diffusion- advection model \cite{Collins_2024} explain the UHE emission from HESS J1825$-$137. \cite{Breuhaus_2021} argued that in electron dominated astrophysical systems like PWN hard electron spectrum naturally occurs in case of inverse Compton dominated cooling. By considering inverse Compton dominated losses, \cite{Breuhaus_2022} also explains the multi-band spectral energy distribution of PeVatron sources LHAASO J2226$+$6057, LHAASO J1825$-$1326, and LHAASO J1908$+0$621 within the framework of PWN. \\

\section{Model Description}
\label{sec:model}

\begin{figure}
	\centering
	
	\includegraphics[width=\linewidth]{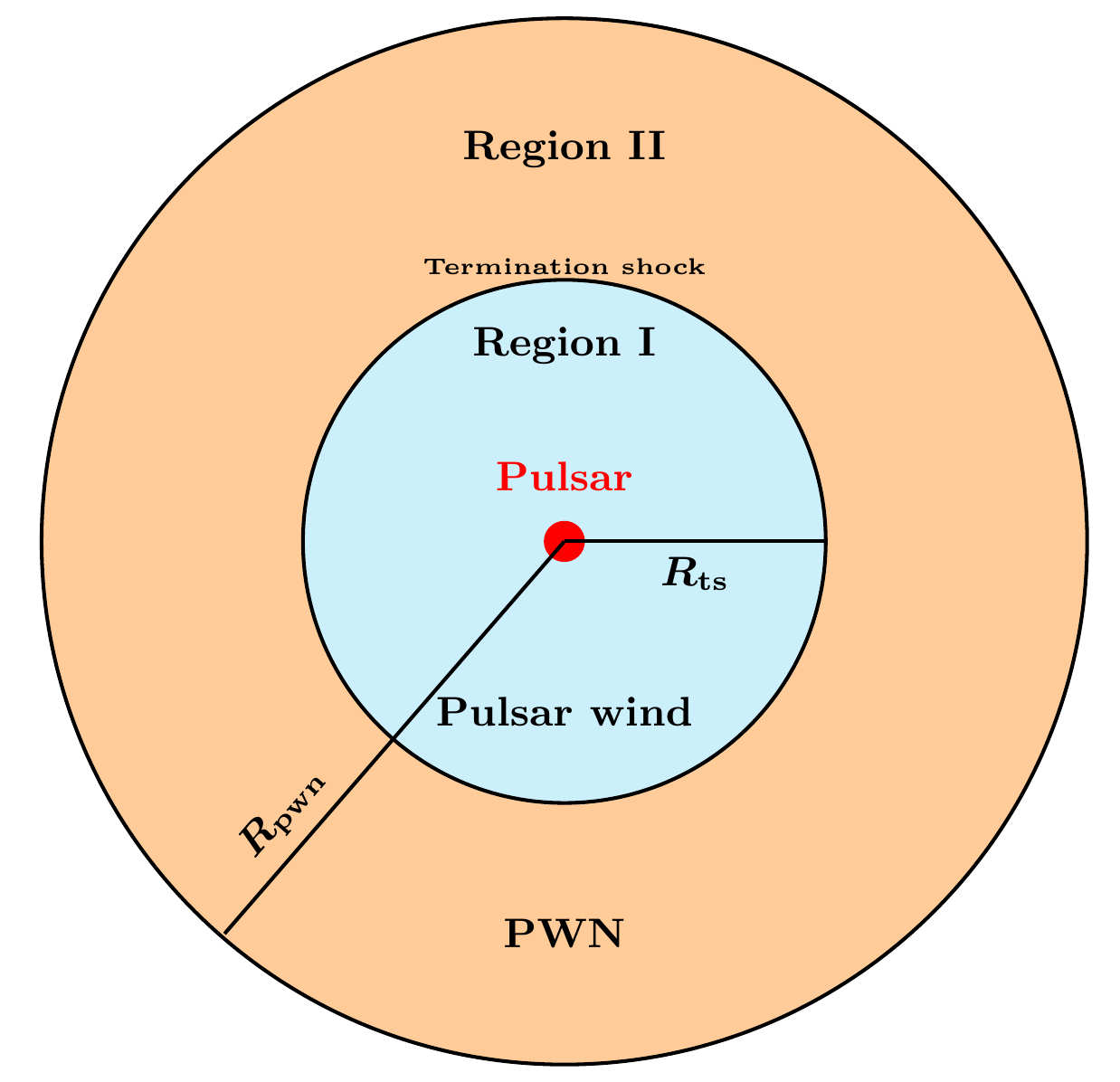}
	\caption{Schematic diagram of Pulsar Wind Nebula model developed in the present work, $R_{pwn}$ is the PWN radius, $R_{ts}$ is the radius of termination shock}
	\label{fig:PWN_diagram}
\end{figure}

In the earlier works authors generally considered a steady injection of energetic electrons into the pulsar wind nebula and those electrons emit radiation and evolve in time. But the spectral form of the injected electrons were assumed and the spectral parameters were determined by fitting the observed data. Here, in the present work we considered the model of pulsar wind as described by \cite{Kennel_Coroniti_1984} and then considered that the electrons were accelerated by the termination shock of the pulsar wind. The shock accelerated electron spectrum is a power-law with exponential cut-off. It is shown that the normalization of the spectrum and the spectral cut-off are completely determined by the magnetization parameter of the pulsar wind. In fact the magnetic field in the post shock region which is essentially the nebular region, can be determined by the magnetization parameter of the pulsar wind.  Therefore, the number of parameters in the model is reduced and the final photon spectrum from the PWN primarily depends on the magnetization parameter. By doing this we attempted to make the model more self consistent and it definitely reduces the parameter space required to describe the broadband spectral energy distribution. The basic model formulation of our work is discussed below. 
\\
\noindent To understand the nature of non-thermal particle in PWN environment and to study the interaction of pulsar wind with nebula, it is important to understand the nature of plasma that pulsar inject to its surrounding. Several theoretical studies argued that pulsar can produce pair plasma in polar cap or outer gap and over all nature of pulsar wind can be explained by considering only (or predominantly) pair plasma as a constituent of pulsar wind \citep{Sturrock_1971,Ruderman_1975, Fawley_1977,Scharlemann_1978,Arons_1979, Arons_1981,Barnard_1982}. In our work, we consider pulsar injects electron positron wind to its surrounding at a ultra-relativistic speed which halts at a distance $R_{ts}$. At $r=R_{ts}$ a stationary shock called termination shock is created. In this work we also assumed spherical symmetry in PWN environment. It should be noted that PWN environment is not really spherically symmetric, e.g. in crab nebula environment there exist a strong north south asymmetry in optical and soft x-ray band, and in radio band it appears as a prolate spheroid. The assumption of spherical symmetry is taken just for mathematical simplicity (for more details see \cite{Kennel_Coroniti_1984}). The schematic diagram of our model is shown in Figure \ref{fig:PWN_diagram}.  \\
In the current work, we divided the Pulsar Wind Nebula (PWN) in to two region, viz. Region I and Region II. Region I started just before the termination shock and end in the immediate downstream. While Region II started at immediate downstream and end at the outer boundary of PWN. In Region II turbulent acceleration, radiative losses, and diffusive escape of particles are considered.

\subsection{\label{sec:Region1}Dynamics of the non-thermal particles in the immediate downstream of the termination shock}

\noindent Pulsar wind is in general ultra-relativistic with a very high Lorentz factor. The rotational energy of the central pulsar decreases and converted into wind luminosity. Following \cite{Kennel_Coroniti_1984}, we assumed that all the spin-down luminosity of the pulsar is converted into wind luminosity.
\begin{equation}
	L_{sd} = 4\pi  R^2_{ts}  \gamma  u  m  c^3 \, (1 \, + \,\sigma)
	\label{Lsd}
\end{equation}
Where $\gamma$ is the upstream bulk Lorentz factor, $u_1$ is the upstream four velocity, $n$ is the proper density, $R_{ts}$ is the radius of the termination shock, and $\sigma$ is the ratio of electro-magnetic energy flux to the particle energy flux, commonly referred as the \textit{magnetization parameter}. 
\begin{equation}
	\sigma = \frac{B^2_1}{4\pi n u_1 \gamma mc^2} 
	\label{sigma}
\end{equation}
where $B_1$ is the upstream magnetic field in the observer frame. The upstream plasma is highly relativistic such that $u_1 \approx \gamma_1$. In pulsar wind magnetic field is dominated by the toroidal component \citep{Goldreich_1969}. So for a $90^{\circ}$ magnetohydrodynamics (MHD) strong shock  Rankine-Hugoniot condition provides the downstream four velocity as 
\begin{multline}
	u_2 \, = \\ \, \frac{8\sigma^2 \, + \, 10\sigma \, + 1 \, + \sqrt{64 \sigma^2(\sigma \, + \, 1)^2 \, + \, 20\sigma(\sigma \, + \, 1)\, + \, 1}}{16(\sigma \, + \, 1)}
	\label{ds_u2}
\end{multline}

\noindent So equation (\ref{ds_u2}) shows that under strong shock approximation (when both sonic and Alfvenic Mach number $>>$ 1) downstream four velocity depends on $\sigma$ only. \cite{Kennel_Coroniti_1984} studied the behavior on downstream medium under high $\sigma$ and low $\sigma$ limit, and shows that downstream parameter of the shock highly depends on $\sigma$ under strong shock approximation and  low sigma shock is more efficient particle accelerator compared to high $\sigma$ shock.\\
The low $\sigma$ limit of equation (\ref{ds_u2}), yields

\begin{align}
	u_2^2 \, = \frac{1 \, + \, 9\sigma}{8} \label{dsu2}\\
	\gamma_2 \, = \, \frac{9 \, + \, 9\sigma}{8} \label{dsgam}\\
	\frac{B_2}{B_1} \, = \, 3(1 \, -\, 4\sigma) \label{dsmf}
\end{align}

It is generally claimed that low $\sigma$ shock is an efficient particle accelerator \citep{Kennel_Coroniti_1984,Ishizaki_2017}, but acceleration of particles in magnetized relativistic perpendicular shock like PWN TS is still an open question. The necessary condition for efficient shock acceleration is that particle should cross the shock front back and forth for several times. This condition is very difficult to achieved in perpendicular relativistic MHD shock \citep{Sironi_2015}, unless the downstream medium is populated by strong turbulence (for example turbulence generated due to the growth of Weibel instability). If turbulence in the immediate downstream is high enough such that $\frac{\delta B}{B} >> 1$, then only particle in the downstream can come back to the upstream for efficient shock acceleration to take place. Such condition can be achieved only in very weakly magnetized medium $\sigma < 10^{-3}$. In such case spectrum of the accelerated particle will be a power with a nearly universal index $\sim$ -2.3 \citep{Amato_2024}. It should be noted that although development of strong turbulence provide the necessary condition for efficient particle acceleration, still this condition is not sufficient for producing high energy particles. Turbulence generated in any astrophysical environment like PWN is small-scale turbulence. Since acceleration time is increases with energy as $\propto E^2$, so such small amplitude turbulence environment fails to produce high energy particles. One possible way to overcome this issue is the existence of large scale turbulence. MHD turbulence generated as a result of corrugation of the termination shock can be a potential candidate for required large amplitude turbulence \citep{Lemoine_2016a}. \cite{Lemoine_2016b} discuss the effect of corrugation and possibility of non-thermal particle acceleration in corrugation of TS.
	
Recently \cite{Kirk_2023} showed that ultra-relativistic shock can be highly efficient even if they fails to produce large amplitude turbulence in the downstream region. \cite{Kirk_2023} argued that if the upstream is highly turbulence dominated, then deflection of particle via regular magnetic field is not prominent, in such case rather than advected with the bulk flow of downstream medium, particle indeed return to upstream and populate a power law with momentum index $\sim$ -4.17. \cite{Dempsey_2007} show that for relativistic shock non-thermal particle spectrum can be expressed as a exponential cut power law as $\sim p^{-\alpha} \; exp \left[-\left(\frac{p}{p_{cut}}\right)^{\beta}\right]$, where $\alpha$ is the spectral index and $p_{cut}$ is the cut-off momentum. The quantity $p_{cut}c$ is the maximum energy of a particle that can be achieved via DSA in perpendicular relativistic shock. Of course, the quantity $p_{cut}$ will depend on pitch-angle in the upstream as well as bulk Lorentz factor of the shock.

\cite{Dempsey_2007} calculated $p_{cut}$ for both non-relativistic shock and relativistic shock, and it can be observed that maximum achievable energy $p_{cut}$ in non-relativistic and relativistic shock is not significantly different. In the present work for the mathematical and computational simplicity we will use the expression of $p_{cut}$ for non-relativistic shock as a first approximation. This will not affect the basic results of the present formalism. \cite{Dempsey_2007} showed that for momentum independent diffusion coefficient, cut-off momentum of non-thermal particle after getting accelerated in non-relativistic MHD shock, can be expressed as

\begin{multline}
	p_{cut} = \frac{2}{\omega_1} 
	\Big( \frac{s^4(2+2\Omega)-s^3(5+11\Omega)+s^2(5+8\Omega-2\chi)}{(s^2(1-\Omega)+6\Omega s-9\Omega)^2} \\
	+ \frac{s(33\Omega+\chi)-36\Omega}{(s^2(1-\Omega)+6\Omega s-9\Omega)^2}\Big) \label{pcuteq}
\end{multline} 

Where the quantity $s$, $\omega$, $\Omega$, and $\chi$ are defined as

\begin{align}
	s \, = \frac{3v_1}{v_1-v_2} \\
	\omega \, =\, \frac{4 \lambda\kappa}{v^2} \label{omeq}\\
	\Omega = \frac{\omega_2}{\omega_1} \label{OMeq}
\end{align}

\begin{equation}
	\chi = (s-3)\sqrt{(s^2+2s^2\Omega+s^2\Omega^2-6s\Omega^2+2s\Omega+9\Omega^2-8\Omega}
	\label{chieq}
\end{equation}

Here the subscript 1 and 2 implies the corresponding quantities in immediate upstream and downstream respectively. $v$ is the magnitude of bulk three velocity. And the constant $\lambda$ is defined as 

\begin{equation}
	\dot{p}_{loss} = \, -a_sB^2p^2 \, = \, - \lambda p^2 \label{lam}
\end{equation}

\noindent In this work, we assume strong shock approximation, hence all the downstream parameters can be expressed in terms of $\sigma$ only. From equation (\ref{dsgam}), downstream 3 velocity can be expressed as 
\begin{equation}
	v_2 = \left(1 - \frac{1}{\gamma^2_2}\right)^{1/2}c \label{eq:v2}
\end{equation}
Since upstream is ultra-relativistic $v_1 \approx c$, so equation (\ref{dsgam}) and (\ref{eq:v2})  gives, compression ratio 
\begin{align}
	\frac{v_1}{v_2} = r = \sqrt{\frac{1+9\sigma}{9+9\sigma}} \label{req}\\
	s = \frac{3r}{r-1}
\end{align}

\begin{figure}
	\centering
	\includegraphics[width=\linewidth]{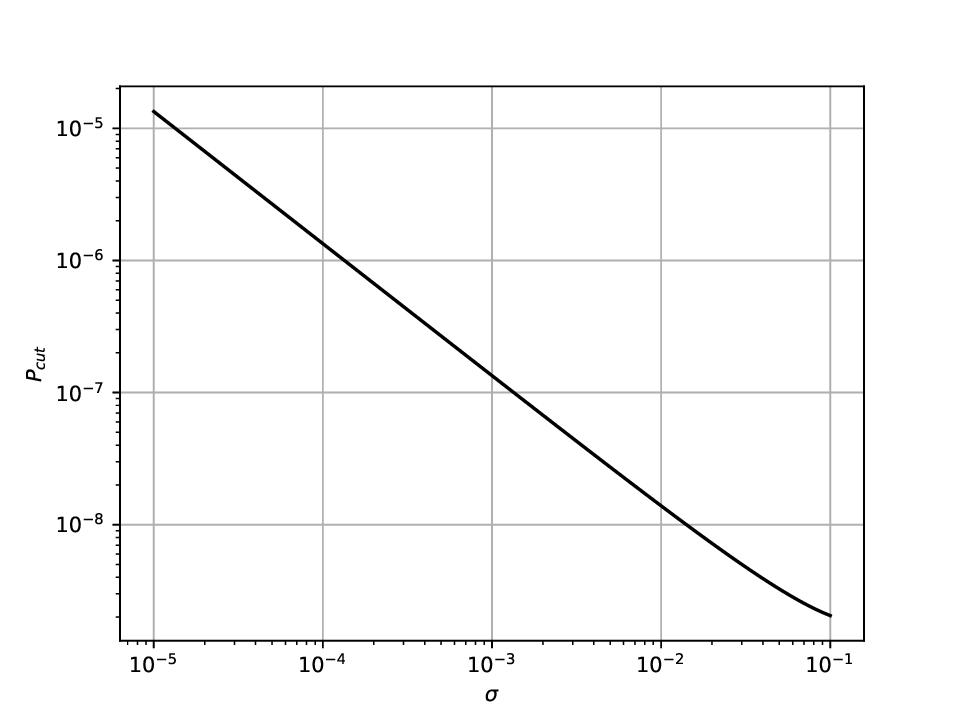}
	\caption{Variation of momentum cut-off ($P_{cut}$) w.r.t. magnetization parameter ($\sigma$) in low $\sigma$ limit}
	\label{fig:Pcut_vs_sigma}
\end{figure}

\noindent From equation (\ref{chieq}) it can be noted that, the quantity $\chi$ depends only on $\Omega$ and $s$. From equation (\ref{OMeq}) and (\ref{omeq}), it can be shown that\\
\begin{equation}
	\Omega = \frac{\omega_2}{\omega_1} = \left(\frac{\lambda_2}{\lambda_1}\right)\left(\frac{\kappa_2}{\kappa_1}\right)\left(\frac{v_1}{v_2}\right)^2
\end{equation}
By considering the dependence of magnetic field with spatial diffusion coefficient $\kappa$ ($\kappa \propto \frac{1}{B}$), and dependence of $\lambda$ with magnetic field $B$ from equation (\ref{lam}), along with equation (\ref{dsmf}) and (\ref{req}), the quantity $\Omega$ can be expressed as 
\begin{equation}
	\Omega = 3(1-4\sigma)\frac{(1+9\sigma)}{(9+9\sigma)}
\end{equation}
From equation (\ref{pcuteq}), it is interesting to note that apart from the term ($\frac{2}{\omega_1}$) all other quantities does not directly depend on the values of physical parameter in the upstream or downstream region, rather they depend on the ratio of physical parameter in upstream to downstream medium. But to calculate the term $\omega_1$ we need the absolute value of $v_1$, $\kappa_1$, and $B_1$. For ultra -relativistic flow, we can consider $v_1 \approx c$, and it is also safe to assume a typical value $3 \times 10^{28} \; cm^2\, s^{-1}$ of diffusion coefficient in PWN environment. Following equation (\ref{Lsd}) and (\ref{sigma}), upstream magnetic field $B_1$ can be constrain with the help of magnetization parameter $\sigma$ as
\begin{equation}
	B_1 = \sqrt{\frac{\sigma L_{sd}}{r_{ts}^2\,c(1+\sigma)}}
	\label{B1eq}
\end{equation}
Also it should be noted that spin-down luminosity of pulsar is a dynamical parameter and it evolve with time.  \cite{Gaensler_2006} showed that for a pulsar with rotational period $p$ and period derivative $\dot{p}$, with initial spin-down luminosity $L_0$, and breaking index $n$, the variation of luminosity with time can be expressed as 
\begin{equation}
	L_{sd}(t) = L_0\left(1+\frac{t}{\tau_0}\right)^{-\frac{n+1}{n-1}}
	\label{pulsar_L_sd_eq}
\end{equation}
Where $\tau_0$ is the initial spin-down time scale of the pulsar, given by $\tau_0 = \tau_c - T_{age}$.  $T_{age}$ is the age of the PWN and $\tau_{c} = \frac{1}{n-1}\frac{p}{\dot{p}}$ is the characteristic age of the pulsar.
Hence it is better to replace $L_{sd}$ by $L_{sd}(t)$ in equation (\ref{B1eq}). Following these aspect, variation of momentum cut-off for different magnetization parameter can be calculated from equation (\ref{pcuteq}). Figure \ref{fig:Pcut_vs_sigma} shows the variation of $p_{cut}$ with $\sigma$ for low $\sigma$ shock.

\subsection{\label{sec:Region2}Effect of turbulent acceleration and radiative losses in Region II: Final Particle spectrum}
\begin{figure}
	\centering
	\includegraphics[width=\linewidth]{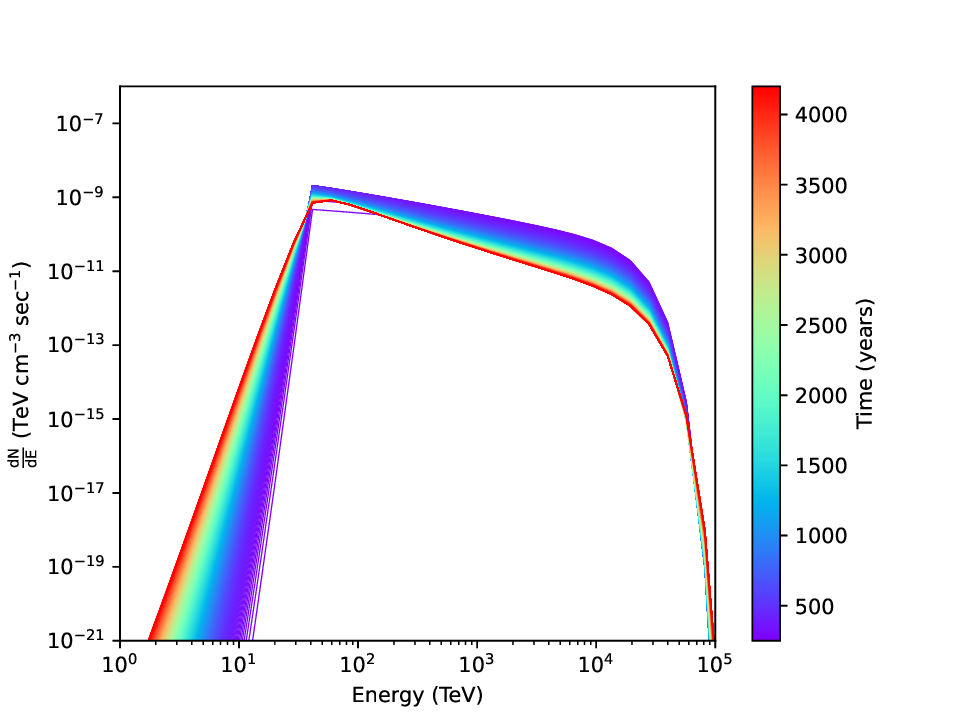}
	\caption{Energy Spectrum evolution of non-thermal electron and positron inside the pulsar wind nebula environment}
	\label{fig:particle_spectrum}
\end{figure}

\noindent After crossing the termination shock the non-thermal particle will enter in the PWN and evolve in its environment. In Region II the evolution will mainly govern by radiative losses, turbulence in the environment and dynamic evolution of the magnetic field of the PWN itself. The equation governing the evolution of non-thermal particle in PWN is given by
\begin{equation}
	\frac{\partial N}{\partial t} = \frac{\partial}{\partial E}\left[ E^2 D_{EE} \frac{\partial}{\partial E} \frac{N}{E^2}\right] + \frac{\partial}{\partial E}\left[ \dot{E}N\right] -\frac{N}{\tau_{esc}} + Q(E,t)
	\label{R2eq}
\end{equation}
Here the term in the left represents the time evolution of the particle spectrum, first term in the right hand side represents the turbulent acceleration of the particle in PWN environment, the 2nd term represents the radiative losses, 3rd term represents the diffusive escape of particle and last term is the injection of non-thermal particle at termination shock of PWN. The significance of each term in equation (\ref{R2eq}) is discussed below.
\begin{itemize}
	\item \textbf{Turbulent acceleration:} The first term on right hand side of equation (\ref{R2eq}) represents turbulent acceleration. Non-thermal particle that crosses the termination shock will enter in PWN environment and interact with the turbulence of the environment and gets accelerated via stochastic acceleration. In this phenomenon, turbulence of the environment plays a crucial role. In our work we consider both Kolmogorov and Kraichnan turbulence in the environment. The signature of the nature of turbulence in the environment will reflect in the momentum diffusion coefficient $D_{pp}$ or in energy space $D_{EE}$. For both the case diffusion coefficient $D_{EE}$ can be express as
	\begin{equation}
		D_{EE} = D_{EE,0}\;\left(\frac{E}{1\;TeV}\right)^{\delta_{EE}}
		\label{Dee}
	\end{equation}
	where for Kolmogorov turbulence $D_{EE,0} \; \sim 6.4 \times 10^{-13} \; TeV^2\,s^{-1}$  and $\delta_{EE}$ = 5/3
	, corresponding values for Kraichnan turbulence are $1.6\times 10^{-12} \; TeV^2 \, s^{-1}$, 3/2 respectively \cite{Lu_2023}.
	
	\item \textbf{Radiative loss:} The second term in equation (\ref{R2eq}) represents the radiative loss term, containing both synchrotron loss and inverse Compton scattering ($\dot{E} \; = \dot{E}_{syn} \, + \, \dot{E}_{IC}$).\\
	The synchrotron loss-rate can be expressed as 
	\begin{equation}
		\dot{E}_{syn} =\frac{4}{3}\sigma_{T}c\left( \frac{B^2_{pwn}(t)}{8\pi}\right)\beta^2 E^2
		\label{syn_loss_rate}
	\end{equation}
	Here $\sigma_T$ is the Thomson cross-section, $\beta$ is the relativistic velocity ($\beta \approx 1$) and $B_{pwn}$ is the magnetic field in the PWN environment. It is important to note that magnetic field in PWN is a dynamical parameter and it evolve with time along with the evolution of other dynamical parameter such as spin down luminosity ($L_{sd}$) and radius of the PWN ($R_{pwn}$). To estimate $B_{pwn}$ let us proceed as follows.\\
	Following equation (\ref{dsmf}) and (\ref{B1eq}) magnetic field in the immediate downstream can be expressed as 
	\begin{equation}
		B_2 = 3(1-4\sigma) \, \sqrt{\frac{\sigma L_{sd}(t)}{r_{ts}^2 c (1+\sigma)}}
		\label{B2_in_sigma}
	\end{equation}
	The MHD conservation law demands magnetic field at a distance $r$ from the pulsar in the nebula to be \cite{Ishizaki_2017}
	\begin{equation}
		B(r) = B_2 \frac{\gamma(r)}{\gamma_2}\frac{u_2}{u(r)}\frac{r_{ts}}{r}
		\label{Br_field}
	\end{equation}
	
	\noindent \cite{Kennel_Coroniti_1984} shows that under small $\sigma$ limit, downstream flow velocity in the nebula at a distance $r$ can be approximate as 
	\begin{equation}
		v(z) = \frac{u(z)}{u_2} \approx \frac{1}{1+\Delta}\left[1+\left(\frac{\bar{z}}{z}\right)^{2/3}\right]
		\label{v(z)eq}
	\end{equation}
	where $z$ is the normalized distance ($z= \frac{r}{r_{ts}}$) and
	\begin{align}
		\bar{z} = \sqrt{\frac{\Delta^3}{(1+\Delta)^2}} \label{zeq}\\
		\Delta = \frac{\left(\frac{u_2}{\sigma}-\frac{1}{2}\right)}{\left(u_2^2+\frac{1}{4}\right)} \label{deltaeq}
	\end{align}
	Substituting equation (\ref{v(z)eq}), (\ref{zeq}),(\ref{deltaeq}),(\ref{B2_in_sigma}), (\ref{pulsar_L_sd_eq}),(\ref{ds_u2}) and (\ref{dsgam}) in equation (\ref{Br_field}) magnetic field in the nebula $B(r)$ can be calculated.\\
	
	In the current study, we are not considering any spatial dependence of particle spectrum. We are interested in evolution and dynamics of the non-thermal particles in presence of second order Fermi acceleration in the turbulence present in nebular flow, and in presence of synchrotron and inverse Compton loss processes. So we can use a volume-averaged magnetic field to serve our purpose. Hence, we define
	\begin{equation}
		B_{pwn}(t) = \sqrt{\frac{\int_{r_{ts}}^{R_{pwn}(t)} B^2(r) d^3r }{\int_{r_{ts}}^{R_{pwn}(t)} d^3 r}}
	\end{equation}
	
	Here, it should be noted that the MHD equations that we have considered is applicable for ideal MHD, which is dissipation free. But in reality due to dissipation and other effect, magnetic field energy reduces from its theoretical value. Hence it is reasonable to consider a correction factor to account for such effects. 
	
	\begin{equation}
		B_{pwn}(t) = \xi \; \sqrt{\frac{\int_{r_{ts}}^{R_{pwn}(t)} B^2(r) d^3r }{\int_{r_{ts}}^{R_{pwn}(t)} d^3r}}
		\label{eq:final mag field}
	\end{equation}
	
	where, $\xi$ is the correction factor. The detail physical explanation of the parameter $\xi$ is discussed in section \ref{sub_sec: Effect of xi}.
	
	Here, radius of the PWN $R_{pwn}(t)$ is also a dynamic parameter and evolve with time as \citep{Chevalier_1992, Blondin_1998}
	\begin{equation}
		R_{pwn}(t) = 1.44 \left( \frac{E_{sn}^3 {L}_0^2}{M_{ej}^5}\right)^{1/10} \; t^{6/5}
	\end{equation}
	Where $E_{sn}$ and $M_{ej}$ is the explosion energy and ejecta mass of the progenitor supernova.
	
	Inverse Compton-loss rate can be expressed as \citep{Blumenthal_1970,Lefa_2012}
	\begin{multline}
		\dot{E}_{IC} = 
		\int\int (E_{\gamma} - E_{ph}) \,\\
		W(E,E_{ph},E_{\gamma}) \, n_{ph}(E_{ph}) \, dE_{ph}dE_{\gamma}
		\label{IC_loss}
	\end{multline}
	Where $E_{ph}$ and $E_{\gamma}$ are the un-scattered and scattered photon energy respectively. $n_{ph}(E_{ph})$ is the distribution of external photon field. In our case external photon field is cosmic microwave background, which is generally represented by a Planckian distribution at 2.7 K temperature.
	\begin{equation}
		n_{ph}(E_{ph}) = \frac{1}{\pi^2 \hbar^3c^3 } \frac{E_{ph}^2}{e^{\frac{E_{ph}}{k_BT}-1}}
		\label{photon_dist}
	\end{equation}
	The function $W(E,E_{\gamma},E_{ph})$ is given by
	\begin{multline}
		W(E,E_{\gamma},E_{ph}) = \\
		\frac{8\pi r_e^2 c}{E \eta}\left[ 2q \, ln \, q+(1-q) \left( 1+2q+ \frac{\eta^2 q^2}{2(1+\eta q)} \right)\right]
	\end{multline}
	with
	\begin{align}
		\eta = \frac{4E_{ph}E}{(m_ec^2)^2} \\
		q = \frac{1}{\eta} \frac{1}{\left( \frac{E}{E_{\gamma}} - 1\right)}
	\end{align}
	
	\item \textbf{Diffusive escape of particles:} The $\mathrm{3^{rd}}$ therm in equation (\ref{R2eq}) represents the escape term, with $\tau_{esc}$ as escape time scale. Since we have considered turbulence in the system, the escape time scale $\tau_{esc}$ will also be determined by the nature of the turbulence present in the environment. In general escape time scale is defined as
	\begin{equation}
		\tau_{esc} = \frac{R^2_{pwn}(t)}{D_{RR}} \label{eq:esc_time}
	\end{equation}
	Where $D_{RR}$ is spatial diffusion coefficient, which is related to momentum diffusion coefficient $D_{EE}$ as \citep{Skilling_1975}
	\begin{equation}
		D_{RR}D_{EE} = \frac{v_A\gamma^2}{9}
		\label{diffusion_coefficient_relation}
	\end{equation}
	$v_A$ is the Alfven velocity in the environment. In analogous to equation (\ref{Dee}) and (\ref{diffusion_coefficient_relation}), spatial diffusion coefficient $D_{RR}$ can be expressed as 
	\begin{equation}
		D_{RR} = D_{RR,0}\left( \frac{E}{1 \; TeV} \right)^{\delta_{RR}}
		\label{Drr}
	\end{equation}
	In Kolmogorov diffusion $D_{RR,0} \, \sim 1.3 \times 10^{26} \; cm^2\, s^{-1}$, and $\delta_{RR} \, \sim 1/3$; while in Kraichnan diffusion corresponding values are $3.5 \times 10^{25} \, cm^2 \, s^{-1}$ and 1/2 respectively.
	
	\item \textbf{Injection spectrum:} The last term of equation (\ref{R2eq}) is the injection term. As discussed in section \ref{sec:Region1}, the non-thermal particle in the immediate downstream will have the spectral distribution of an exponential cut power law in momentum space as $f(p) \propto p^{-\alpha} \, exp\, \left(-\left(\frac{p}{p_{cut}}\right)^{\beta}\right)$. Number spectrum in immediate downstream in energy space can be calculated as by $4\pi p^2 f(p)$ and substituting $E \approx pc$ for non-thermal particles. With this number spectrum in the immediate downstream become 
	\begin{equation}
		N(E) \propto E^{2-\alpha} \, exp\left(-\left(\frac{E}{E_{cut}}\right)^{\beta}\right) 
		\label{N(E)}
	\end{equation}
	Where $E_{cut} = p_{cut}c$ is the cut-off energy. With this injection spectrum in Region II become,
	\begin{equation}
		Q(E) = \frac{N(E)}{\tau^{I}_{esc}}
		\label{Q(E)}
	\end{equation}
	Where $\tau^I_{esc}$ is the escape time -scale in Region I. For momentum independent diffusion coefficient, escape time $\tau^I_{esc}$ will be a constant in energy space. With equation (\ref{N(E)}) and (\ref{Q(E)}), the injection spectrum become
	\begin{equation}
		Q(E) = Q_0E^{2-\alpha} \, exp\left( - \left( \frac{E}{E_{cut}}\right)^{\beta}\right)
		\label{Q(E)ECPL}
	\end{equation}
	Both the proportionately constant and $\tau^I_{esc}$ is absorbed in $Q_0$. The normalization $Q_0$ can be estimated from
	\begin{equation}
		\int^{E_{max}}_{E_{min}} E\, Q(E) \, dE \, = \, \eta_p \, L_{sd}(t) \frac{1}{\int^{R_{pwn}(t)}_{r_{ts}} d^3r} \label{Q0eq}
	\end{equation}
	Here $\eta_p$ is the fraction of particle luminosity to the total spin-down luminosity of the pulsar \citep{Gelfand_2009}.
	\begin{align}
		\eta_p = \frac{L_p}{L_{sd}} \label{eta_p}\\
		\eta_B = \frac{L_B}{L_{sd}}
	\end{align}
	with $\eta_p + \eta_B =1$. From equation (\ref{Lsd}) and (\ref{eta_p}) $\eta_p$ can be estimated as 
	\begin{equation}
		\eta_p = \frac{1}{1+\sigma} \label{etap_sigma}
	\end{equation}
	Substituting equation (\ref{etap_sigma}) and (\ref{Q(E)ECPL}) in (\ref{Q0eq}), normalization $Q_0$ can be estimated as
	\begin{multline}
		Q_0(t) = \\
		\frac{1}{\int^{R_{pwn}(t)}_{r_{ts}} d^3r}  \frac{\left(\frac{1}{1+\sigma} \right)L_{sd}(t)}{\int^{E_{max}}_{E_{min}}E^{3-\alpha} \, exp\left(-\left( \frac{E}{E_{cut}} \right)^{\beta} \right) \, dE} 
	\end{multline}
\end{itemize}
With these term we solve equation (\ref{R2eq}) numerically by following Chang Cooper algorithm \citep{Chang_1970, Park_numerical_1996}, by considering no-flux boundary condition (For more details see \cite{Park_analytical_1995, Park_numerical_1996}). The time evolution of particle spectrum in PWN environment is shown in Figure \ref{fig:particle_spectrum}.

\begin{figure}[ht!]
	\centering
	\includegraphics[width=\linewidth]{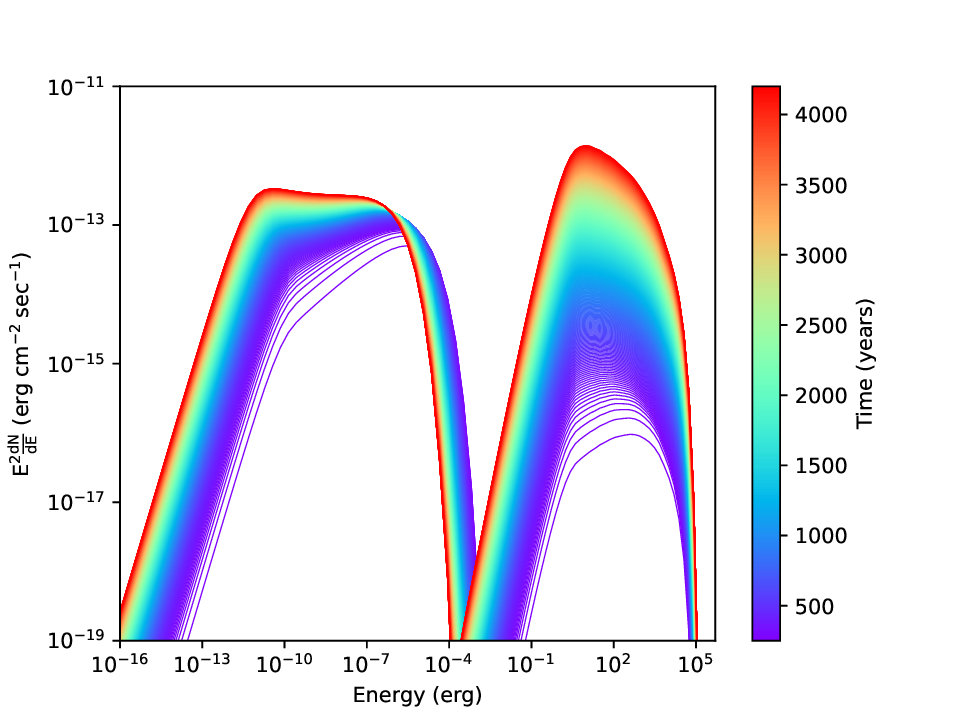}
	\caption{Broadband energy spectrum of photons emitted by the non-thermal particles (electron and positron) in PWN environment}
	\label{fig:sed}
\end{figure}

\noindent The photon spectrum radiated by the non-thermal particle is calculated by following \cite{Blumenthal_1970,Lefa_2012} (see equation (11) of \cite{Celli_2020}, equation (2.49) of \cite{Blumenthal_1970}, equation (2) of \cite{Lefa_2012}). Figure \ref{fig:sed} shows the broadband spectral energy distribution of photons emitted by the non-thermal particles in PWN environment through synchrotron radiation and inverse Compton scattering with cosmic microwave background. 

\section{\label{sec:Results}Results and Discussion}

\noindent In this work we investigate the particle acceleration and radiation emission mechanism in PWN by following a MHD approach. One important aspect of this formalism is that in this formalism most of the model parameter can be expressed in terms of the magnetization parameter $\sigma$. As discussed in section \ref{sec:Region1}, the cut-off energy $E_{cut}$ of the injection spectrum can be determined by $\sigma$ only. Apart from $\sigma$, the shock radius ($R_{ts}$) and nature of turbulence also plays a crucial role in determining particle spectrum as well as spectral energy distribution of emitted photons. In this section we have discussed the effect of each parameter on the particle spectrum of the non-thermal electrons and the broadband spectral energy distribution of photons emitted by the non-thermal electrons.

\subsection{Effect of magnetization parameter}
\noindent \cite{Kennel_Coroniti_1984} shows that under strong shock approximation downstream quantities depends on $\sigma$ only. Based on \cite{Kennel_Coroniti_1984} results current model allow us to examine the dependence on the parameter $\sigma$ on particle spectrum of the accelerated electrons and broadband spectral energy distribution of photons.
\begin{figure*}
	\includegraphics[width=\linewidth]{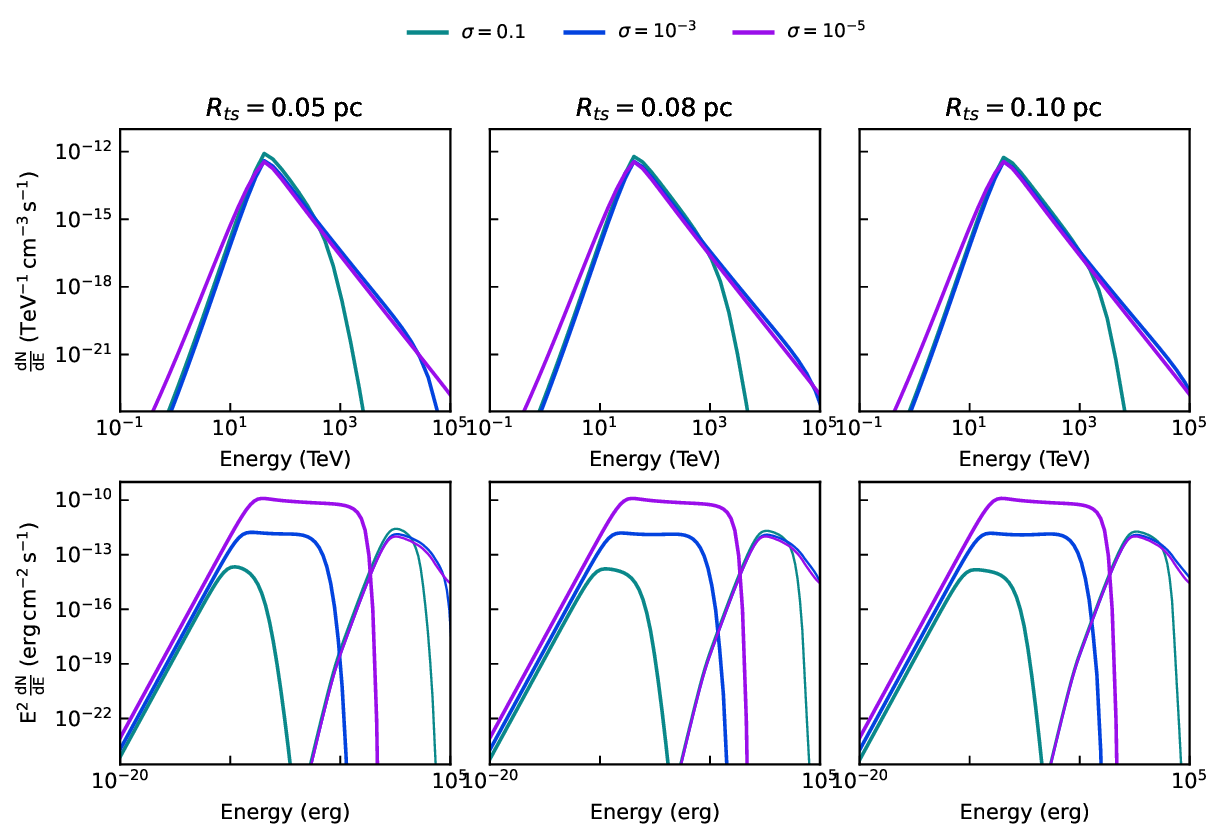}
	\caption{Effect of magnetization parameter $\sigma$ on particle spectrum and spectral energy distribution of photons. Upper panel shows the variation of particle spectrum in PWN with $\sigma$ at three different shock radius $R_{ts}$ = 0.05 pc, 0.08 pc and 0.1 pc respectively. Different value of $\sigma$ is shown by different color in the figure. Lower panel represents the corresponding spectral energy distribution of the emitted photons.}
	\label{fig:sigma_variation}
\end{figure*}

\noindent Figure \ref{fig:sigma_variation} shows the variation of particle spectrum of electrons and corresponding SED of emitted photons with $\sigma$ at three different values of shock radius (0.05 pc, 0.08 pc, 0.1 pc). Since we are working in small $\sigma$ approximation, highest value of $\sigma$ in our model can be 0.1 \citep{Kennel_Coroniti_1984}. We computed the particle spectrum and SED
for 3 different values of $\sigma$ ($10^{-5}, 10^{-3}, 10^{-1}$).

\noindent As already discussed, $\sigma$ effect the injection spectrum and thereby effecting final spectral distribution of photons. With increase in $\sigma$, the region become magnetic flux dominated and hence causes more losses due to synchrotron radiation. Which limits on the maximum achievable energy of the non-thermal particles (see figure \ref{fig:Pcut_vs_sigma}). This shows that low $\sigma$ wind are more favorable candidate for particle acceleration up to ultra high energy (UHE).

\noindent The parameter $\sigma$ also controls the normalization of injection spectra. The relation between $\sigma$ and $\eta_p$ as shown in equation (\ref{etap_sigma}) implies that with increase in $\sigma$, $\eta_p$ decreases. Which implies that fraction of pulsar spin down luminosity converted into particle luminosity is reduced, this effectively reduces the number of particle injected in the termination shock and consequently reduces the emitted photon flux.

\noindent Since with increase in $\sigma$, $\eta_p$ decreases or $\eta_B$ increases, so the ratio of synchrotron flux to IC flux also effected by change in $\sigma$ (see lower panel of Fig \ref{fig:sigma_variation}). 

\subsection{Effect of shock radius}
\begin{figure*}[ht!]
	\includegraphics[width=\linewidth]{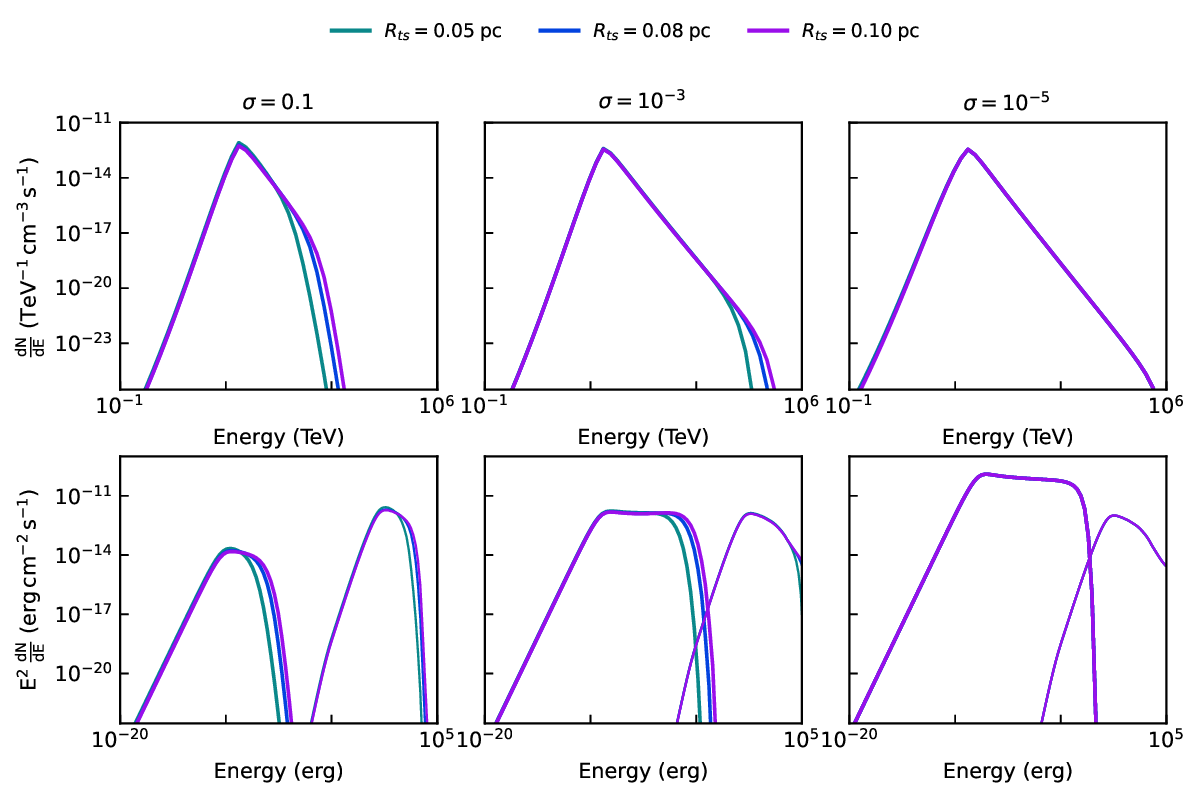}
	\caption{Effect of position of shock radius ($R_{ts}$) on particle spectrum and spectral energy distribution of photons. Upper panel shows the variation of particle spectrum in PWN with $R_{ts}$ at three different values of $\sigma$ = $10^{-1}, \, 10^{-3}$ and $10^{-5}$ respectively. Different value of $R_{ts}$ is shown by different color in the figure. Lower panel represents the corresponding spectral energy distribution of the emitted photons.}
	\label{fig:Rts_variation}
\end{figure*}

\noindent The location of the pulsar wind termination shock sets the physical environment in which particles are injected and accelerated. \cite{Ishizaki_2017} discussed the effect of shock radius on the particle spectrum. As the magnetic field in the immediate upstream region depends on the location of termination shock, the particle spectrum injected into the nebular region and the final photon spectrum will eventually depend on the position of the termination shock. \\
Figure~\ref{fig:Rts_variation} shows the dependence of the particle spectrum on the termination shock radius, $R_{\mathrm{ts}}$, for three representative values of $\sigma$. The figure demonstrates that the maximum achievable particle energy increases for larger shock radii. This is because a smaller $R_{\mathrm{ts}}$ corresponds to a stronger upstream magnetic field (see equation ~(\ref{B1eq})) causing the particles to lose energy by synchrotron process faster than the situation when $R_{ts}$ is relatively large.

The influence of the shock radius is most pronounced at relatively large values of $\sigma$. As $\sigma$ decreases, the flow becomes increasingly particle dominated and the sensitivity to $R_{\mathrm{ts}}$ is reduced. Accordingly, in Fig.~\ref{fig:Rts_variation} the effect is clearly visible for $\sigma = 0.1$, becomes restricted to the vicinity of the high-energy cutoff for $\sigma = 10^{-3}$, and is negligible for $\sigma = 10^{-5}$, where the weakly magnetized environment causes the spectra for different shock radii to nearly overlap.

\subsection{Effect of turbulence} 
In the relativistic plasma inside PWN, the presence of MHD turbulence plays the most crucial role in the spatial and momentum diffusion process of the charged particles. Hence, the evolution of the particle spectra within the source is highly dependent on the nature of turbulence present within it. In the present work, we investigated the effect of two types of MHD turbulence - Kolmogorov turbulence and Kraichnan turbulence in the evolution of the electron distribution in Region-II. The electron energy distribution and the corresponding broadband photon energy distribution for the two turbulence scenarios are shown in Figure~\ref{fig:variation_with_xi_turb}. In both turbulence cases, the electron spectra exhibited a power-law behavior up to an energy $\rm \sim 10^4 ~TeV$, with an exponential cut-off thereafter, dominated by radiative cooling. At energies lower than the injection energy of the electrons ($\rm \sim 1 TeV$), a steeper particle spectrum is obtained in the case of Kolmogorov turbulence as compared to the Kraichnan turbulence. At the lower energy, the escape time-scale for the Kolmogorov turbulence is found to be always lower for the given standard parameters, hence the particle escape is more in this case in comparison to Kraichnan turbulence. In PWN, the effect of the interaction of these two types of turbulence in accelerating the electrons is more prominent for a lower magnetic field i.e a lower $\xi$ value.  In case of a lower value of $\xi$, the effect of radiative cooling of the accelerated particles is comparatively lower, which leads to a harder particle spectrum for both the turbulence scenario, as shown in Figure~\ref{fig:variation_with_xi_turb}. A detailed discussion of the effect of $\xi$ is summarized in section \ref{sub_sec: Effect of xi}. As the Kolmogorov turbulence is more efficient in accelerating particles to higher energies~\citep{Becker_2006, Ghosal_2024}, the particle spectrum in the case of Kolmogorov turbulence becomes harder as compared to the Kraichnan turbulence. The effect of the two  turbulence scenarios on the particle spectrum is reflected in the broadband spectral energy distribution of photons as shown in the  lower panel of Figure~\ref{fig:variation_with_xi_turb}. In case of low $\xi$ ($10^{-2}$ to $10^{-4}$) the synchrotron spectrum become harder around x-ray band in case of Kolmogorov turbulence, also the IC peak shifted to higher frequency.

\subsection{\label{sub_sec: Effect of xi}Effect of magnetic field correction factor}
\noindent The quantity $\xi$ in equation (\ref{eq:final mag field}) is the magnetic field correction factor. Physically, the correction factor $\xi$ represents the loss of magnetic field energy in PWN environment. In astrophysical environment like PWN, magnetic field energy can be deviated from ideal MHD results due to several reasons. 

\noindent In this work, we use \citet{Kennel_Coroniti_1984} MHD approach to study the behavior of non-thermal particles in PWN environment. It is important to note that this formalism is applicable to ideal MHD and hence applicable only to non-dissipative system. In reality, in most of the astrophysical environment, this assumption does not hold, and a significant loss in magnetic field energy can be observed due to dissipation loss. So deviation from ideal MHD results in decrease in magnetic field strength, and hence a correction factor is required.

\noindent Apart from dissipation, we have also consider the presence of strong turbulence in PWN environment. Consideration of turbulence itself indicates deviation from ideal MHD scenario. The turbulence in PWN environment interact with the non-thermal particles injected by the pulsar, and as a result of wave particle interaction non-thermal particles gain energy from the wave and get accelerated. The particle accelerated in the PWN at a cost of loss of magnetic turbulence energy of the environment. This non-linear wave-particle interaction also lead to decrease in magnetic field energy of the PWN.

\begin{figure}
	\centering
	\includegraphics[width=\linewidth]{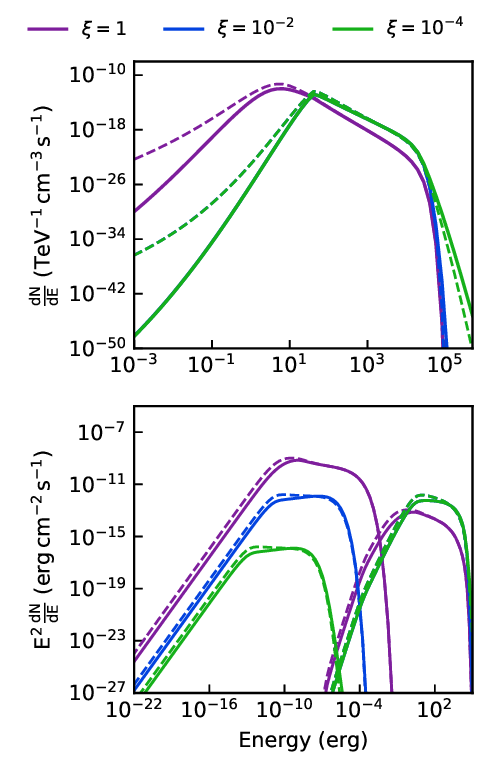}
	\caption{Effect of magnetic field correction factor $\xi$ on particle spectrum of non-thermal particles and spectral energy distribution of photons for both Kolmogorov and Kraichnan turbulence. Different value of $\xi$ is represented by different color in the figure. Solid line and dashed line represent Kolmogorov and Kraichnan turbulence respectively.}
	\label{fig:variation_with_xi_turb}
\end{figure}

\noindent In this work, we consider a correction factor $\xi$ to account for all such losses in PWN environment. It is important to note that $\xi$ has a significant effect on the final particle spectrum as well as on the broadband spectral energy distribution of photons. Fig \ref{fig:variation_with_xi_turb} shows the variation of particle spectrum and sed with magnetic field correction factor $\xi$. Generally the maximum achievable energy in any astrophysical kinetic process can be obtained by equating acceleration rate with radiation loss rate (such as synchrotron loss). Since, synchrotron loss rate highly depends on the magnetic field strength in the system, so maximum achievable energy of non-thermal particle also depends on $\xi$, i.e. increases with increase in $\xi$. Apart from this, the quantity $\xi$ shows major effect on broadband sed as well. As the synchrotron radiation directly depends on magnetic field strength, so with decrease in $\xi$, corresponding flux of synchrotron radiation also decreases as a result of which radio to x-ray flux of the source gets reduced. With decrease in $\xi$, synchrotron peak is also sifted towards lower energy.

It is important to note that effect of $\xi$ on particle spectrum is much prominent for lower values of $\xi$. For example, in the upper panel of fig. \ref{fig:variation_with_xi_turb}, blue curve (corresponding to $\xi = 10^{-2}$) nearly overlaps with the green curve (corresponding to $\xi = 10^{-4}$), except at high energy. As at $\xi=10^{-4}$ magnetic field is significantly reduces, so maximum achievable energy of the particle increases as indicated by acceleration rate- loss rate balance condition. Although effect of $\xi$ is suppressed in the particle spectrum (except near cut-off energy rangy), the variation in the synchrotron spectrum of the emitted photons with respect to $\xi$ is clearly visible. As $\xi$ directly related to the magnetic field, so effect of different $\xi$ can be explained as a consequence of different synchrotron loss rate. On the other hand as $\xi$ has no effect on inverse Compton scattering loss rate so IC spectrum of blue and the green curve is again overlapping like the parent non-thermal particle population.

\section{\label{sec:application}Application of the model} 
\noindent We have applied our model in two different LHAASO detected PeVatron which have spatial association with PWN viz. 1LHAASO J1929+1846u, 1LHAASO J1848-0001u. These sources were primarily chosen because they were detected by LHAASO with high TS value, they were also detected by HESS and HAWC and finally the data were available. We constrain the model parameter to explain the observed UHE emission from the sources.

\subsection{1LHAASO J1929+1846u}
\noindent The sources 1LHAASO J1929+1846 is one of the UHE source in LHAASO 1st catalog detected with a TS of 416.2 which have spatial association with PWN \citep{Cao_2024}. A PWN G54.1+0.3 is located within $r_{39}$ radius of the UHE source powered by a pulsar PSR J1930+1852. The source G54.1+0.3 is firstly detected at 4.75 GHz by \cite{Reich_1984}. VLA, OSRT and IRAS observation reveal that the source G54.1+0.3 is a pleronic system like crab nebula \citep{Velusamy_1988}. The first x-ray image and spectrum of the source is reported by \cite{Lu_2001,Lu_2002} with the ASCA SIS, ROSAT PSPC and GIS observation. Later \cite{Camilo_2002} also discovered the pulsar PSR 1930+1852 with a period of 136 ms through radio observation. Recently \cite{Abeysekara_2018} confirmed the association of the PWN G54.1+0.3 with UHE $\gamma$-ray source VER J1930+188 and 2HWC J1930+188. Recent observation of LHAASO shows that the PWN G54.1+0.3 can emmit UHE $\gamma$-ray and hence a galactic PeVatron \citep{Cao_2024}. Here we tried to explain the UHE emission of the source 1LHAASO J1929+1846u by considering the PWN G54.1+0.3 as an acceleration site. The model parameter used in this work is reported in table \ref{tab:model_parameter_J1929}.

\begin{figure}
	\centering
	\includegraphics[width=\linewidth]{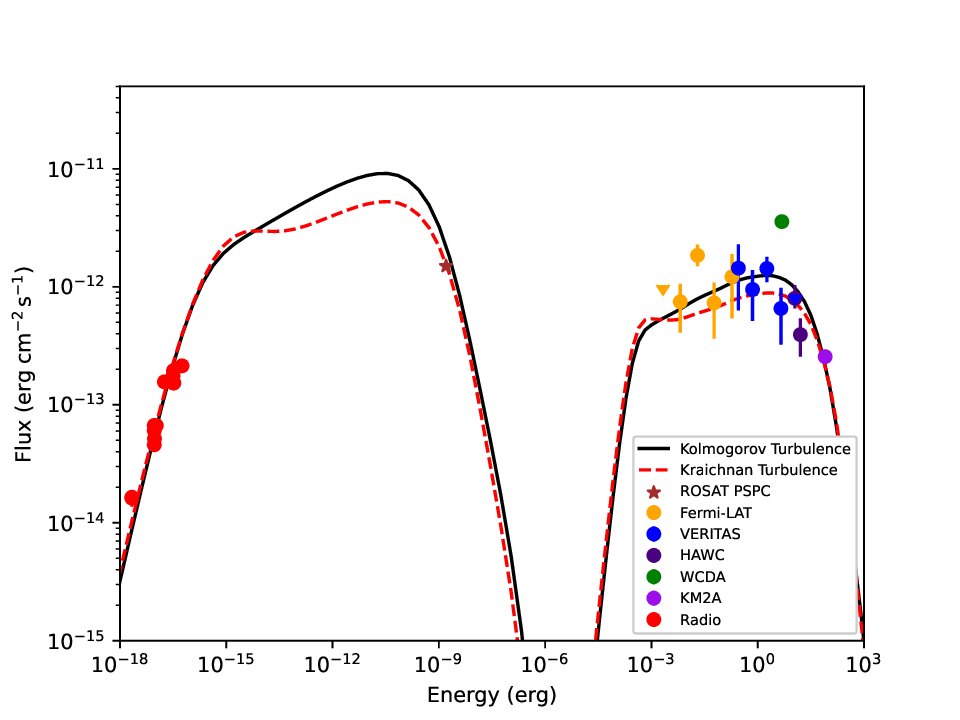}
	\caption{Broadband spectral energy distribution of the source 1LHAASO J1929+1846u. Black solid line, and red dashed line represent the model flux of the source considering Kolmogorov and Kraichnan turbulence respectively. Different flux points used in the SED are taken from \cite{Cao_2024} (LHAASO KM2A and LHAASO WCDA), \cite{Albert_2020, Albert_2023} (HAWC), \cite{Abeysekara_2018} (VERITAS), \cite{Xia_2023} (Fermi LAT), \cite{Lu_2001} ROSAT PSPC, \cite{Xia_2023} and references therein (Radio).}
	\label{fig:J1929_combined}
\end{figure}

\begin{table*}[t]
	\centering
	\caption{Parameters used to fit the broadband spectral energy distributions of the PeVatron sources 1LHAASO J1929+1846u considering both Kolmogorov and Kraichnan turbulence.}
	\begin{tabular}{llcc}
		\toprule
		& \textbf{Physical Parameter} & \textbf{Kolmogorov turbulence} & \textbf{Kraichnan turbulence} \\
		\midrule
		\multirow{4}{*}{Model Parameter} 
		& Magnetization parameter, $\sigma$ & 0.08 & 0.08 \\
		& Radius of termination shock, $R_{\mathrm{ts}}$ & 0.1 pc & 0.1 pc \\
		& Momentum index of injection spectrum, $\alpha$ & 4.4 & 4.4 \\
		& Magnetic field correction factor, $\xi$ & 0.1 & 0.075\\
		\midrule
		\multirow{4}{*}{Turbrlence parameter} & Energy diffusion coefficient, $D_{\mathrm{EE,0}}$ & $6.4 \times 10^{-13}$ $\mathrm{TeV^2 \, s^{-1}}$ & $1.6 \times 10^{-12}$ $\mathrm{TeV^2 \, s^{-1}}$\\
		& Turbulence index in energy space, $\delta_{EE}$ & 5/3 & 3/2 \\
		& Spatial diffusion coefficient, $D_{\mathrm{RR,0}}$ & $1.3 \times 10^{26}$ $\mathrm{cm^2 \, s^{-1}}$ & $3.5 \times 10^{25}$ $\mathrm{cm^2 \, s^{-1}}$\\
		& Turbulence index in position space, $\delta_{RR}$ & 1/3 & 1/2 \\
		\midrule
		\multirow{2}{*}{SN parameter} 
		& SN explosion energy, $E_{\mathrm{SN}}$ & $5 \times 10^{51}$ erg & $5 \times 10^{51}$ erg \\
		& Ejecta mass, $M_{\mathrm{ej}}$ & 3 $M_\odot$ & 3 $M_\odot$ \\
		\midrule
		\multirow{5}{*}{Pulsar parameter} 
		& Spin-down luminosity, $L(t)$ & $1.18 \times 10^{37} \; \mathrm{erg\,s^{-1}}$ & $1.18 \times 10^{37} \; \mathrm{erg\,s^{-1}}$\\
		& Initial spin-down luminosity, $L_0$ & $2.2 \times 10^{38} \; \mathrm{erg\,s^{-1}}$ & $1.7 \times 10^{38} \; \mathrm{erg\,s^{-1}}$ \\
		& Period, $p$ & 136 ms & 136 ms \\
		& Period derivative, $\dot{p}$ & $7.5 \times 10^{-13} \; \mathrm{s\,s^{-1}}$ & $7.5 \times 10^{-13} \; \mathrm{s\,s^{-1}}$ \\
		& Braking index, $n$ & 3 & 3 \\
		\midrule
		\multirow{2}{*}{PWN parameter} 
		& Age of the PWN, $T_{\mathrm{age}}$ & 2140 yr & 1970 yr \\
		& Distance, $d$ & 6 kpc & 6 kpc \\
		\midrule
		\multirow{3}{*}{Electron energy parameter} 
		& Minimum electron energy, $E_{\mathrm{min}}$ & 100 eV & 100 eV \\
		& Maximum electron energy, $E_{\mathrm{max}}$ & 50 PeV & 50 PeV \\
		& Injection energy, $E_{\mathrm{inj}}$ & 6 GeV & 3 GeV \\
		\bottomrule
	\end{tabular}
	
	\label{tab:model_parameter_J1929}
\end{table*}

\subsection{1LHAASO J1848-000u}
\noindent The source 1LHAASO J1848-000u is first detected by LHAASO with a TS of 655.4 by KM2A \citep{Cao_2024}. The position of the source is reported to be RA = 282.74$^{\circ}$, and DEC = -0.02$^{\circ}$. The source is identified as PWN, around the pulsar PSR J1849-0001. The UHE $\gamma$-ray source is spatially coincide withe the soft-gamma ray source IGR J18490-000, which is at an angular distance $0.08^{\circ}$ away from the LHAASO detected center. The source was previously also detected by HESS in HESS galactic plane survey \citep{HESS_Collaboration_b_2018} and named the source as HESS J1849-000. The pulsar, PWN system of the source is first resolved by XMM-Newton \citep{Terrier_2008}. Using Rossi X-ray Timing Explorer (RXTE) observation \cite{Gotthelf_2011} identify the pulsar PSR J1849-0001 and estimate the parameters of the pulsar such as period (p), period derivative ($\dot{p}$) and current spin-down luminosity L(t) as \citep{Manchester_2005}\\
p = 38.52 ms, $\dot{p}$ = $ 14.14 \times 10^{-15} \; s\,s^{-1} $, L(t) = $9.77 \times 10^{36} \; erg \, s^{-1}$.

\noindent In this work we modeled the observed UHE emission and broadband spectral energy distribution of photons from the source 1LHAASO J1848-000u using our two zone model. The model parameter used to explain the observed SED are summarized in table \ref{tab:model_parameter_J1848}. 

\begin{figure}
	\centering
	\includegraphics[width=\linewidth]{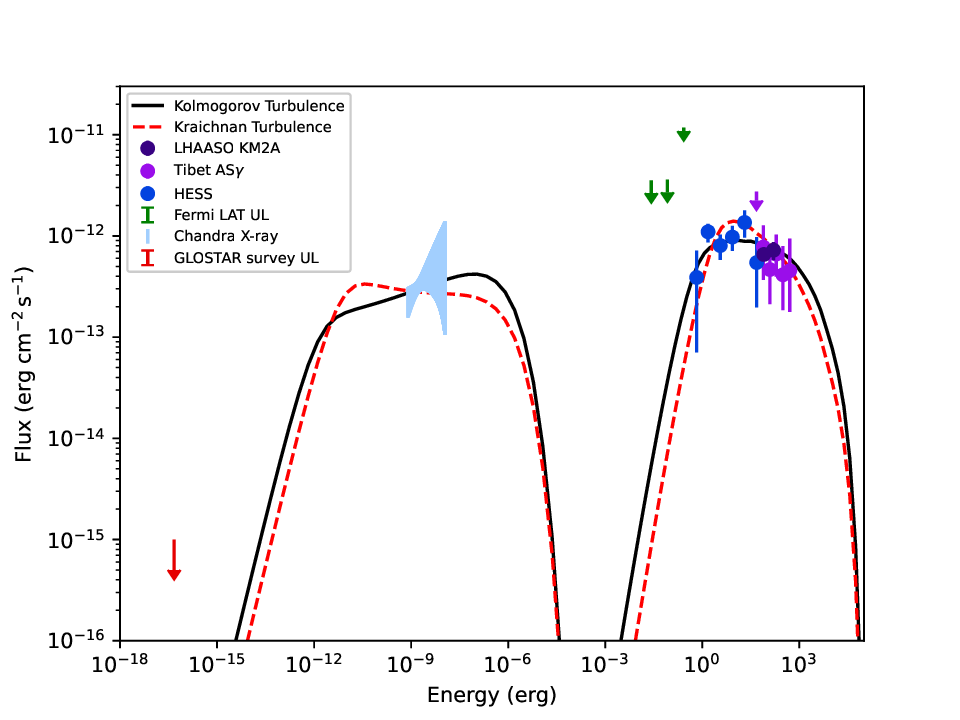}
	\caption{Broadband spectral energy distribution of the source 1LHAASO J1848-000u. Black solid line and red dashed line represent the model flux of the source considering Kolmogorov and Kraichnan turbulence respectively. Different flux points used in the SED are taken from \cite{Cao_2024, Cao_2021} (LHAASO KM2A), \cite{Amenomori_2023} (Tibet AS$\gamma$), \cite{HESS_Collaboration_b_2018} (HESS), \cite{Acero_2013} (Fermi LAT),  \cite{Gagnon_2024} (Nustar), \cite{Brunthaler_2021} (GLOSTAR survey).}
	\label{fig:J1848_combined}
\end{figure}

\noindent Fig \ref{fig:J1848_combined} shows the broadband spectral energy distribution of the source 1LHAASO J1848-000u. With the set of parameter given in table \ref{tab:model_parameter_J1848}, the UHE emission can be explained. It is important to note that the broadband SED of the source can be explained by considering both Kolmogorov and Kraichnan diffusion.

We would like to clarify that a preliminary validation of the model is presented
using the multi wavelength data for two PWN. The comparison with the data
is qualitative and no formal statistical fitting is performed at this stage.
Table \ref{tab:model_parameter_J1929} and \ref{tab:model_parameter_J1848} list the corresponding physical parameters for the two sources.

\begin{table*}[t]
	\centering
	\caption{Parameters used to fit the broadband spectral energy distributions of the PeVatron sources 1LHAASO J1848-000u considering both Kolmogorov and Kraichnan turbulence.}
	\begin{tabular}{llcc}
		\toprule
		& \textbf{Physical Parameter} & \textbf{Kolmogorov turbulence} & \textbf{Kraichnan turbulence} \\
		\midrule
		\multirow{4}{*}{Model Parameter} 
		& Magnetization parameter, $\sigma$ & 0.005 & 0.005 \\
		& Radius of termination shock, $R_{\mathrm{ts}}$ & 0.1 pc & 0.1 pc \\
		& Momentum index of injection spectrum, $\alpha$ & 4.17 & 4.17 \\
		& Magnetic field correction factor, $\xi$ & 0.0038 & 0.0035\\
		\midrule
		\multirow{4}{*}{Turbulence parameter} & Energy diffusion coefficient, $D_{\mathrm{EE,0}}$ & $6.4 \times 10^{-13}$ $\mathrm{TeV^2 \, s^{-1}}$ & $1.6 \times 10^{-12}$ $\mathrm{TeV^2 \, s^{-1}}$\\
		& Turbulence index in energy space, $\delta_{EE}$ & 5/3 & 3/2 \\
		& Spatial diffusion coefficient, $D_{\mathrm{RR,0}}$ & $1.3 \times 10^{26}$ $\mathrm{cm^2 \, s^{-1}}$ & $3.5 \times 10^{25}$ $\mathrm{cm^2 \, s^{-1}}$\\
		& Turbulence index in position space, $\delta_{RR}$ & 1/3 & 1/2 \\
		\midrule
		\multirow{2}{*}{SN parameter} 
		& SN explosion energy, $E_{\mathrm{SN}}$ & $1.5 \times 10^{51}$ erg & $1.5 \times 10^{51}$ erg \\
		& Ejecta mass, $M_{\mathrm{ej}}$ & 4 $M_\odot$ & 4 $M_\odot$ \\
		\midrule
		\multirow{5}{*}{Pulsar parameter} 
		& Spin-down luminosity, $L(t)$ & $9.77 \times 10^{36} \; \mathrm{erg\,s^{-1}}$ & $9.77 \times 10^{36} \; \mathrm{erg\,s^{-1}}$\\
		& Initial spin-down luminosity, $L_0$ & $1.25 \times 10^{37} \; \mathrm{erg\,s^{-1}}$ & $1.2 \times 10^{37} \; \mathrm{erg\,s^{-1}}$ \\
		& Period, $p$ & 38.52 ms & 38.52 ms \\
		& Period derivative, $\dot{p}$ & $1.41 \times 10^{-14} \; \mathrm{s\,s^{-1}}$ & $1.14 \times 10^{-14} \; \mathrm{s\,s^{-1}}$ \\
		& Braking index, $n$ & 3 & 3 \\
		\midrule
		\multirow{2}{*}{PWN parameter} 
		& Age of the PWN, $T_{\mathrm{age}}$ & 5000 yr & 4200 yr \\
		& Distance, $d$ & 7 kpc & 7 kpc \\
		\midrule
		\multirow{3}{*}{Electron energy parameter} 
		& Minimum electron energy, $E_{\mathrm{min}}$ & 100 eV & 100 eV \\
		& Maximum electron energy, $E_{\mathrm{max}}$ & 50 PeV & 50 PeV \\
		& Injection energy, $E_{\mathrm{inj}}$ & 20 TeV & 40 TeV \\
		\bottomrule
	\end{tabular}
	
	\label{tab:model_parameter_J1848}
\end{table*}

\section{\label{sec:conclusion}Conclusion} 

\noindent In this work, we have developed a two-zone leptonic model to study particle acceleration and broadband emission from pulsar wind nebulae detected at PeV energies by LHAASO. Here we couple the magnetohydrodynamic description of the pulsar wind and the non-thermal particle population in the nebula in a self consistent manner. In contrast to previous phenomenological models, the particle injection spectrum at the termination shock is constrained directly from the magnetization parameter of the pulsar wind, providing a direct physical link between the wind properties and the observed radiation.

By doing so, the magnetic field strength in the immediate upstream, immediate downstream and nebular region, injection spectrum of particle to the pulsar wind nebula and subsequent evolution of particles in the nebular region can be primarily determined from the \textit{magnetization parameter} $\sigma$. As a result, this formulation significantly reduces the effective parameter space. During the particle evolution, we also consider the stochastic acceleration of particles in PWN.

As a test of the model, we apply it to the LHAASO sources 1LHAASO~J1929+1846u and 1LHAASO~J1848$-$000u and successfully reproduce their broadband spectral energy distributions. Our results identify that in low-$\sigma$ environments, the efficient turbulent acceleration of the particles could be the key physical condition for the formation of PeVatrons in pulsar wind nebulae.

These results will help to understand the origin of the ultra-high-energy photons from the LHAASO detected PeVatron. Further studies of several LHAASO detected PeVatrons will be investigated with the present model in the future.

\section*{Acknowledgments}
We sincerely thank the anonymous reviewer for their valuable suggestions, which have significantly improved the quality of the manuscript.




\bibliographystyle{elsarticle-harv} 
\bibliography{main}






\end{document}